\documentclass[journal, a4paper, twoside]{IEEEtran}
\usepackage{amsmath,amssymb,amsfonts,amsthm}
\usepackage{algorithmic}
\usepackage{graphicx,pgfplots}
\usepackage{textcomp,multirow}
\usepackage{caption,subcaption}
\usepackage{bbm}

\newtheorem{theorem}{\textbf{Theorem}}

\newtheorem*{prof}{\textbf{Proof}}
\pgfplotsset{compat=1.17} % added because of a warning
\newcommand{\Kappa}{\mathcal{K}}
\title{Multi-Access Cache-Aided Multi-User Private Information Retrieval}
\author{Kanishak Vaidya, and B Sundar Rajan \\
Department of Electrical Communication Engineering, IISc Bangalore, India \\
E-mail: \{kanishakv, bsrajan\}@iisc.ac.in \vspace{-3mm}
}

\begin{document}
    \maketitle
    \thispagestyle{empty}

\begin{abstract}
	
	We consider the problem of multi-access cache-aided multi-user Private Information Retrieval (MuPIR). In this problem, several files are replicated across multiple servers. There are $K$ users and $C$ cache nodes. Each user can access $L$ cache nodes, and every cache node can be accessed by several users. Each user wants to retrieve one file from the servers, but the users don’t want the servers to know their demands. Before the users decide their respective demands, servers will fill the cache nodes from the content of the files. Users will then request their desired files from the servers. Servers will perform coded transmissions, and all the users should get their desired files from these transmissions and the content placed in the caches they are accessing. It is required that any individual server should not get any information about the demands of the users. This problem is an extension of the dedicated cache-aided MuPIR problem, which itself generalizes the widely studied single user PIR setup. In this paper, we propose a MuPIR scheme which utilizes a multi-access setup of the coded caching problem. The presented scheme is order optimal when $K = \binom{C}{L}$ users. We also characterize the rate of the scheme for the special case of cyclic wraparound multi-access setup, where $C=K$ and each user access $L$ consecutive cache nodes in cyclic wraparound fashion.
\end{abstract}

\section{Introduction}
    The problem of Private Information Retrieval (PIR) first described in~\cite{Chor95PIR} deals with privately retrieving data from distributed servers. A user wishes to retrieve one file amongst a set of files stored across the servers. But the servers should not know the identity of the desired file. A PIR scheme that minimizes the download cost for the user is described in~\cite{Sun17PIR}. After that the PIR problem has been solved for various other settings \cite{Sun18cPIR,Lin19wPIR,Chen20siPIR}.

    Currently, PIR is being studied with another content delivery scenario called coded caching. Described in~\cite{MAli13CodedCaching}, in coded caching there are multiple users equipped with user cache and one server storing some files. During off peak hours, users fill their caches and then during peak network traffic hours demand files from the server. The server will perform coded transmissions such that a single transmission can benefit multiple users simultaneously. After receiving the transmissions, users will be able to decode their demanded files with the help of the content stored in cache. Recently, in~\cite{Ming21CaMuPIR} a cache-aided PIR strategy is described where multiple users, each having access to dedicated caches, want to privately recover files from non-colluding servers. An order optimal strategy is described that combine coding benefits of PIR in~\cite{Sun17PIR} and coded caching~\cite{MAli13CodedCaching}.

    In this paper, we use a variation of coded caching known as multi-access coded caching in PIR\@. In multi mccess coded caching users don't have access to dedicated caches, instead there are helper cache nodes, which are accessed by the users. One helper cache can be accessed by multiple users and user can access multiple caches. We will be using the multi-access setup described in~\cite{pooja21MANMA} which generalize the Maddah-Ali Niesen coded caching scheme \cite{MAli13CodedCaching}.

       \textit{Notations}: For integers $m$ and $n$, $[m:n]$ is the set of integers $\{ m, m+1, \cdots, n\}$. $[N]$ is same as $[1:N]$. For a set $\mathcal{S}$ of size $|\mathcal{S}|$ and an integer $N \leq |\mathcal{S}|$, $\binom{\mathcal{S}}{N}$ denotes the set of all subsets of $\mathcal{S}$ of size $N$. For the set $\{a_n | n \in [N]\}$ and $\mathcal{N} \subseteq [N]$, $a_{\mathcal{N}}$ denotes the set $\{a_n | n \in \mathcal{N}\}$. % $m \% n \triangleq m \mod{n}$ if $m$ is not integer multiple of $n$ and $m \% n = n$ if $m$ is a multiple of $n$. Fors set $\mathcal{N} \subseteq [N]$, $\mathcal{N} \% n \triangleq \{ i \% n | i \in \mathcal{N}\}$.

 We will first briefly describe the single user PIR of~\cite{Sun17PIR} and the multi-access coded caching setup of~\cite{pooja21MANMA} in the following two subsections.

\subsection{Private Information Retrieval~\cite{Sun17PIR}}
    In Private Information Retrieval (PIR) there is one user and a set of $N$ files $\mathcal{W} = {\{W_n\}}_{n = 1}^N$ replicated across $S$ non-colluding servers. The user wants to retrieve one out of $N$ files, say the file $W_{\theta} , {\theta} \in \{1 , 2 , \hdots , N\}$, but doesn't want the servers to know the identity of the file. In other words, the user wants to hide the index ${\theta}$ from the servers. In order to retrieve this desired file privately, the user generates $S$ queries ${\{Q_s^{\theta}\}}_{s = 1}^S$ and sends the query $Q_s^{\theta}$ to server $s$. After receiving their respective queries, the servers construct answers which are function of the query they got and the files they have. The server $s$ constructs the answer $A_s^{\theta}(\mathcal{W})$ and sends it to the user. After receiving the answers from all $S$ servers, the user should be able to decode its desired file. Privacy and correctness conditions are formally stated as follows: For privacy we need that 
\[
    I({\theta} ; Q_s^{\theta}) = 0 , \forall s \in \{1 , \hdots , S\},
\]
and for correctness
\[
    H(W_{\theta} | {\theta} , A_{1}^{\theta}(\mathcal{W}) \hdots A_{S}^{\theta}(\mathcal{W}) , Q_{1}^{\theta} \hdots Q_{S}^{\theta}) = 0.  
\]
The rate of PIR is a parameter that describes the download cost for the user (or the transmission cost for the servers)   defined as
\[
    R_{PIR} = \frac{\sum_{s = 1}^{S}(H(A_s^{\theta}(\mathcal{W})))}{H(W_{\theta})}.
\]
A rate optimal scheme is provided in~\cite{Sun17PIR} with optimal rate $R^*_{PIR}$ given by
\[
    R^*_{PIR} = 1 + \frac{1}{S} + \frac{1}{S^2} + \hdots + \frac{1}{S^{N-1}}.
\]
\subsection{Multi-Access Coded caching~\cite{pooja21MANMA}}
    In a multi-access coded caching scenario, a server stores $N$ files ${\{W_n\}}_{n \in [N]}$ each of unit size. There are $K$ users, connected via an error free shared link to the server. There are $C$ helper cache each capable of storing $M$ files. Each user has access to $L$ out of $C$ helper caches and there is only one user corresponding to every choice of $L$ out of $C$ caches resulting in $K={C \choose L}.$ Let $\mathcal{Z}_k \subset [C]$ be the indices of helper caches that the user $k$ has access to. The system operates in two phases.

    \textit{Placement Phase}: In this phase, all $C$ helper cache are filled without the knowledge of future demands of the users. Let $Z_{c}$ denote the content stored in the helper cache $c \in [C]$. The content stored in the helper caches is a function of the files $W_{[1:N]}$ and all the servers know the content stored at each of the helper cache.

    \textit{Delivery Phase}: In this phase, each user wishes to retrieve a file from the server. User $k$ will choose $d_k \in [N]$ and wish to retrieve $W_{d_k}$. Users will convey their demands to the server and server will perform coded transmissions such that each user gets the demanded file using the transmissions and the cache content they have access to.

    The goal in this setup is to design a placement and delivery scheme such that the transmissions from the server is minimized. A placement and delivery scheme is given in~\cite{pooja21MANMA} for this setup. This placement and delivery scheme generalize the well known Maddah-Ali Niesen (MAN) scheme for dedicated cache system, i.e., for $L = 1$ and $C = K$ the scheme in \cite{pooja21MANMA} becomes the MAN scheme. Every file is divided into $\binom{C}{t}$ non overlapping subfiles of equal size and the server transmits $\binom{C}{L+t}$ such subfiles where $t = \frac{CM}{N} \in \mathbb{Z}$. The rate achieved in this scheme is $\frac{\binom{C}{L+t}}{\binom{C}{t}}$.

    In~\cite{Brunero21flCMAC} this rate is shown to be optimal under the assumption of uncoded cache placement. We denote this rate by 
    \begin{equation}\label{no_PIR_rate}
        R^*_{nPIR}(t) = \frac{\binom{C}{L+t}}{\binom{C}{t}}        
    \end{equation}
    where the subscript \textit{nPIR} is to indicate that the rate is for the multi-access setup of \cite{pooja21MANMA} with \textit{no PIR} constraints.

\subsection{Our Contributions}

 Multi-User Private Information Retrieval (MuPIR) problem has been studied with coded caching setup  in \cite{Ming21CaMuPIR}. In MuPIR there are multiple users, each equipped with a cache and multiple servers. Each user wants to retrieve a file from the servers but doesn't want the servers to know the user demand.

    In this paper, we  develop a PIR scheme in which multiple users aided with multi-access cache nodes privately retrieve data from distributed servers. We will consider the multi-access setup of~\cite{pooja21MANMA} but with multiple non-colluding servers with all the messages replicated across all the servers. Each server is connected to all the users via broadcast links. 
    
    The contributions and the outline of this paper follow.
    \begin{itemize}
        \item In Section~\ref{system_model} we describe the system model in detail and provide formal description of privacy and correctness constraints.
        \item In Section~\ref{main_results} we present our results.
            \begin{itemize}
                \item In Theorem~\ref{Point_rates} we present an achievable rate for the multiuser PIR problem with multi-access setup of~\cite{pooja21MANMA}. The achievable scheme is given in Section~\ref{scheme}.
                \item In Theorem~\ref{order_optimality} we show that the rate achieved in Theorem~\ref{Point_rates} is order optimal within a multiplicative factor of 2 assuming uncoded storage.
                \item In Theorem~\ref{Cyclic_wraparound} we give an achievable rate for the special case of cyclic wraparound cache access systems widely studied \cite{HKD,ReK,TDTR,CWLZC,SaR,SPE}. The achievable scheme is given in Section~\ref{cyclic_scheme}.
            \end{itemize}
        \item We have compared per user rate \cite{KMRTcom} of our setup with dedicated cache setup of~\cite{Ming21CaMuPIR} in the following four settings.
         
            \begin{itemize}
                \item With same number of caches and same cahe sizes in both dedicated cache and multi-access system.
                \item With same number of caches and equal amount of memory accessed by both users in both settings.
                \item With equal number of users equal number of cache and same memory capacity of cache in both settings.
                \item Comparison with the dedicated cache multi-access setup with cyclic wraparound cache access.
            \end{itemize}
        \item Sections ~\ref{scheme} and ~\ref{cyclic_scheme} contain  the proofs of privacy  and achievable rates mentioned in Theorem~\ref{Point_rates} and Theorem~\ref{Cyclic_wraparound} respectively. 
    \end{itemize}

\section{System Model}\label{system_model}
    \begin{figure}
        \centering
        \includegraphics[width = 0.45\textwidth]{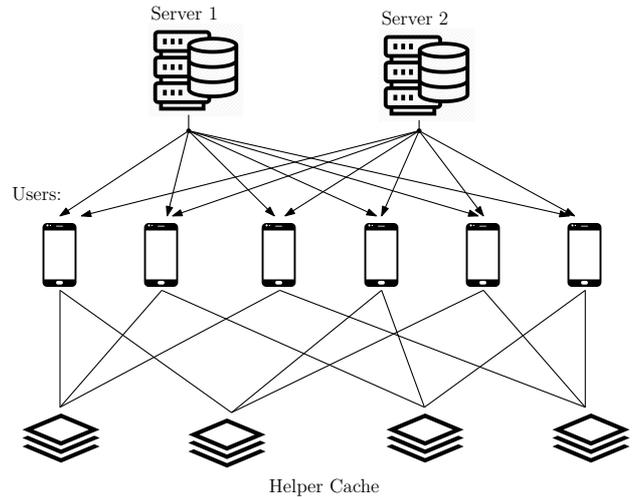}
        \caption{Multi Access coded caching setup with two servers, four helper cache and six users. Each user have access to two helper cache}%
        \label{MAsystem}
    \end{figure}
    There are $K$ users and $N$ files, ${\{W_n\}}_{n \in [N]}$ replicated across $S \geq 2$ servers. Each file is of unit size. There are $C$ cache nodes each capable of storing $M$ files. Every user is connected to $L < C$ cache nodes through an infinite capacity link. Let $\mathcal{L}_k$ be the set of indices of cache nodes that user $k$ is connected to. The system operates in two phases.

    \textit{Placement Phase}: In this phase, all $C$ cache nodes are filled. Let $Z_c$ denote the contents stored in cache $c \in [C]$. $Z_c$ is a function of files $W_{[1:N]}$ and all the servers know the contents stored at each of the helper cache.

    \textit{Private Delivery Phase}: In this phase, each user wish to retrieve a file from the servers. User $k$ will choose $d_k \in [N]$ and wish to retrieve $W_{d_k}$ privately from the servers. Let $\textbf{d} = (d_1 , d_2 , \hdots , d_K)$ be the demand vector. In order to retrieve their desired files from the servers, users will cooperatively generate $S$ queries $Q_s^{\textbf{d}} , s \in [S]$ based on their demands and the content stored in helper cache. Then the query $Q_s^{\textbf{d}}$ will be sent to server $s , \forall s \in [S]$. These queries are constructed such that they don't disclose the demand vector $\textbf{d}$ to any of the server. After receiving their respective queries, server $s , \forall s \in [S]$, will broadcast $A_s^{\textbf{d}}$ to the users which is a function of the query $Q_s^{\textbf{d}}$ and $W_{[1:N]}$. After receiving $A_{[S]}^{\textbf{d}}$, user $k$, $\forall k \in [K]$, will decode $W_{d_k}$ with help of the caches it has access to.
    
    Formally we can say that, in order to preserve the privacy of demands of the users, the following condition needs to be satisfied: 
    \[
        I(\textbf{d} ; Q_s^\textbf{d} , Z_{[1:C]}) = 0 , \forall s \in [S].
    \]
    This condition, known as the privacy condition, ensures that none of the servers have any information about user demands. The condition
    \[
        H(W_{d_k} | \textbf{d} , Z_{\mathcal{L}_k} , A_{[S]}^{\textbf{d}}) = 0 , \forall k \in [K]
    \]
    known as the correctness condition, ensures that users will have no ambiguity about their demanded file.

    The rate $R$ is defined to be the amount of data that has to be transmitted by all the servers in order to satisfy the user demand, and is given by
    \[
        R = \sum_{s = 1}^S H(A_s^{\textbf{d}}).
    \]
    Our goal is to design  placement and private delivery phases jointly that satisfy the privacy and correctness conditions and minimizes the rate. 

\section{Main Results}\label{main_results}
In this section we present the main results of the paper. For a given multi-access cache-aided MuPIR problem we give a scheme to achieve the rate described in Theorem~\ref{Point_rates}. The scheme is described in Section~\ref{scheme}.
\begin{theorem}\label{Point_rates}
    For the multi-access coded caching setup, with $S$ servers, $N$ files, $C$ helper caches and $K = \binom{C}{L}$ users, where each user is accessing a unique set of $L$ helper cache and each cache can store $M$ files and $t = \frac{CM}{N}$ is an integer, the users can retrieve their required file privately i.e.\ without reveling their demand to any of the servers, with rate
    \[
        R(t) = \frac{\binom{C}{t+L}}{\binom{C}{t}} \big( 1 + \frac{1}{S} + \hdots + \frac{1}{S^{N-1}} \big)
    \]
\end{theorem}
\begin{prof}
    A scheme, along with the proof of privacy for the scheme, is given in Section~\ref{general_scheme} that achieve the rate stated above. \hfill \qedsymbol
\end{prof}

Theorem~\ref{Point_rates} gives an achievable rate in multi-access setup where cache memory $M$ is an integer multiple of $N/C$. For intermediate memory points the lower convex envelope of points
\[
    {\{(t , R(t))\}}_{t \in [0:C]}
\]
can be achieved using memory sharing.

Next we show that the rate achieved in Theorem~\ref{Point_rates} is order optimal within a factor of $2$ under the  assumption of uncoded cache placement.

\begin{theorem}\label{order_optimality}
    Under the assumption of uncoded cache placement, the rate achieved in Theorem~\ref{Point_rates} is less than or equal to twice the optimal worst-case rate $R^*(t)$ i.e.
    \[
        R(t) \leq 2 R^*(t).
    \]
\end{theorem}
\begin{prof}
    The optimal rate achieved in multi-access coded caching without PIR constraint can only be as high as the optimal rate achievable in multi-access coded caching setup with PIR constraint. So using \eqref{no_PIR_rate} we have,
    \begin{align*}
        R^*(t) &\geq R^*_{nPIR}(t) \\
        \implies \frac{R^*(t)}{R(t)} &\geq \frac{R^*_{nPIR}(t)}{R(t)} \\
        \implies \frac{R(t)}{R^*(t)} &\leq \big( 1 + \frac{1}{S} + \hdots + \frac{1}{S^{N-1}} \big).
    \end{align*}
    As $\big( 1 + \frac{1}{S} + \hdots + \frac{1}{S^{N-1}} \big) \leq 2$ for all $S \geq 2$ we have
    \[
        R(t) \leq 2 R^*(t).
    \] \hfill \qedsymbol
\end{prof}

The rate achieved in Theorem~\ref{Point_rates} corresponds to the case when the number of users is $\binom{C}{L}$. But if the number of users is less than $\binom{C}{L}$ then a lower rate may be achieved, provided each user is accessing a unique set of $L$ caches. We consider the case when there are $C$ users and users are accessing caches in cyclic wraparound manner i.e.\ user $k$ is accessing caches indexed by $\{ k , k+1 , \hdots k+L-1 \}$ where sum are modulo $C$ except $k+l = C$ if $k+l$ is multiple of $C$. This cyclic wraparound coded caching has been widely studied \cite{HKD,ReK,TDTR,CWLZC,SaR,SPE}. Before stating the rate for this setup, for integers $m \leq k < n$, we define $\mbox{cyc}(n , k , m)$ to be the number of $k$ sized subsets of $n$ distinguishable elements arranged in a circle, such that there is atleast one set of $m$ consecutive elements amongst those $k$ elements. An expression for $cyc(n , k , m)$ is given in \eqref{cyceqn}.
    \begin{figure}[h]
        \centering
        \includegraphics[width = 0.45\textwidth]{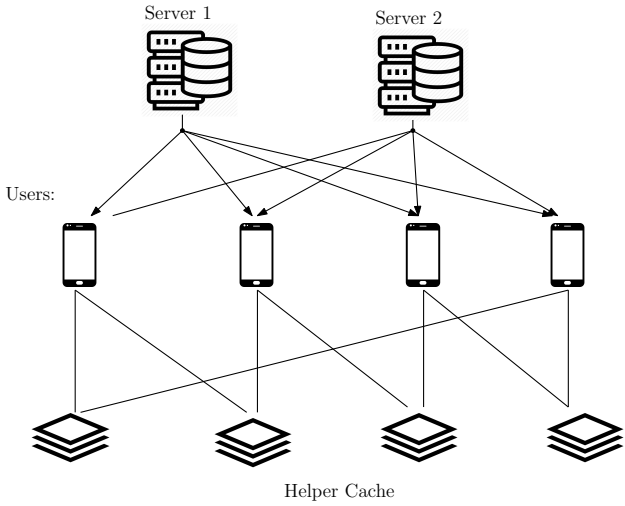}
        \caption{Multi Access coded caching setup with two servers, four helper cache and four users accessing caches in cyclic wraparound manner. Each user have access to two helper cache}%
        \label{Cyclic-WA-system}
    \end{figure}
%\[
%    cyc(n , k , m) = |\{ \mathcal{K} \in \binom{[0:n-1]}{k} : \{a , a+1 , \hdots , a+(m-1)\} \subseteq \mathcal{K} \mbox{ for some } a \in [0:C-1]  \}|
%\]

\begin{figure*}
    \begin{align}
        cyc(n , k , m) &= \sum_{r=1}^k \Big(\binom{n-k}{r} + \binom{n-k-1}{r-1}\Big)\sum_{l=1}^r {(-1)}^{l-1}\binom{r}{l}\binom{k - l(m-1) -1}{r-1} \notag \\
        &+ \sum_{r=3}^k \binom{n-k-1}{r-2}\Big( \sum_{l=2}^{m-1}(l-1) \sum_{j = 1}^{r-2} {(-1)}^{j-1} \binom{r-2}{j} \binom{k - l - j(m-1) -1}{r-3} \label{cyceqn} \\
        &+ \sum_{l=m}^{k} (l-1) \binom{k-l-1}{r-3} \Big) + k-1 \notag
    \end{align}%
    \hrulefill
\end{figure*}

\begin{theorem}\label{Cyclic_wraparound}
    For the multi-access coded caching setup, with $S$ servers, $N$ files, $C$ helper caches and $K = C$ users, where each user is accessing $L$ helper cache in cyclic wraparound manner and each cache can store $M$ files and $t = \frac{CM}{N}$ is an integer, the users can retrieve their required file privately i.e.\ without reveling their demand to any of the servers, with rate
    \[
        R(t) = \min\Big\{ \frac{C - t}{t + 1} , \frac{cyc(C , t+L , L)}{\binom{C}{t}} \Big\} \big( 1 + \frac{1}{S} + \hdots + \frac{1}{S^{N-1}} \big).
    \]
\end{theorem}
\begin{prof}
    In Section~\ref{cyclic_scheme} we present an achievable scheme that achieve the rate stated above for the cyclic wraparound cache access setup. \hfill \qedsymbol
\end{prof}

\subsection{Comparison with dedicated cache setup of~\cite{Ming21CaMuPIR}}
    In this subsection we compare our scheme with the \textit{product design} given in~\cite{Ming21CaMuPIR}. First we  give a brief summary of the product design.

    In dedicated cache setup of~\cite{Ming21CaMuPIR} there are $N$ files ${\{W_n\}}_{n \in [N]}$ replicated across $S$ servers. There are $K$ users, each equipped with dedicated cache capable of storing $M$ files. Users want to retrieve their desired files from the servers. The system works in two phases. In \textit{Placement Phase}, the cache of each user is filled with some content. This cache content is a  function of the files stored across the servers, and is independent of the future demands of the users. Then in \textit{Private Delivery Phase}, each user will choose a file independently and wish to retrieve its respective file form the servers privately. For that, the users will cooperatively generate $S$ queries, and send them to the servers. The servers after receiving their respective queries, will respond with answers. Specifically, server $s$ will broadcast $A_s$ for all $s \in [S]$. In~\cite{Ming21CaMuPIR} an achievable scheme called \textit{Product Design} is given. The product design  achieves the rate $R_{PD}$ given by 
    \begin{align*}
        R_{PD} &= \frac{K - t}{t + 1} (1 + \frac{1}{S} + \hdots + \frac{1}{S^{N-1}}) \\
        &\mbox{where} \\
        t &= \frac{KM}{N}
    \end{align*}

    Note that, the rate achieved by the product design is same as the rate achieved in Theorem~\ref{Point_rates} for the special case of $L = 1$ i.e., when every user is accessing only one cache.

    In coded caching system, the number of cache nodes and the storage capacity of each cache node are crucial parameters. In dedicated cache systems, number of cache nodes and number of users supported in the networks are same. In contrast, multi-access coded caching systems can support more number of users for the same number of cache nodes. So, in multi-access systems, even if the number of transmissions are more than that of dedicated cache system, it is possible that one transmission is beneficial to more number of users. So we will be considering the parameter \textit{per user rate} or \textit{rate per user} i.e.\ $\frac{R}{K}$ for comparing two systems. This was introduced in \cite{KMRTcom}. For distinction, quantities related to dedicated cache system will have subscript $DC$ and multi-access setup quantities will have subscript $MA$. We compare our scheme with the product design in the following four settings with $2$ servers and $3$ files.  
    \begin{itemize}
        \item Both the dedicated cache system and the multi-access system have the same number of caches i.e.\ $C$ cache nodes in both settings and the cache size is also same in both settings i.e.\ $M_{DC} = M_{MA} = M$. In this case there will be $C$ users in the dedicated cache setup and $\binom{C}{L}$ users in the multi-access setup.
        \item Both the dedicated cache system and the multi-access system have the same number of caches i.e. $C$ cache nodes in both settings but each user is accessing the same amount of memory. As users in the multi-access system are accessing $L$ cache nodes each of size $M_{MA}$ and in the dedicated cache system each user is accessing only one cache of size $M_{DC}$ we will set $M_{DC} = L \times M_{MA}$. In this case also, there will be $C$ users in the dedicated cache setup and $\binom{C}{L}$ users in the multi-access setup.
        \item The number of users in both systems are same i.e.\ $K_{MA} = K_{DC}$ and total system memory is also same. Considering $C$ cache nodes in multi-access system we have $K_{MA} = \binom{C}{L} = K_{DC}$. And as the number of cache nodes in the dedicated cache system is same as the number of users there will be $\binom{C}{L}$ cache nodes in the dedicated cache system. For the same total memory in both settings, we want $M_{DC} \times \binom{C}{L} = M_{MA} \times C$.
        \item The number of cache nodes and the number of users are same in both settings i.e.\ $K_{MA} = K_{DC}$ and $C_{MA} = C_{DC}$. Moreover, as $K_{DC} = C_{DC}$ we will consider $C$ cache and $C$ users in both settings. For this, we consider cyclic wraparound cache access i.e.\ user $k$  will access cache nodes $\{k , k+1 , \cdots , k+L-1\}$ where sum is modulo $C$ except $k+l = C$ is $k+l$ is multiple of $C$. We will also consider identical cache sizes in both the multi-access and dedicated cache setup $M_{DC} = M_{MA}$. 
    \end{itemize}

    Also note that the parameter $t = \frac{CM}{N}$ in multi-access setup and $t = \frac{KM}{N}$ in dedicated cache setup denote how many times entire set of $N$ files can be replicated across the cache. For instance, if $t = 2$ then cache nodes are capable of storing $2N$ units. Also, total memory of system is $tN$ units, which is equal to $KM$ for dedicated cache system and $CM$ for multi-access setup. For dedicated cache system, number of cache nodes is always equal to the number of users $K$.

    \subsubsection{Same number of cache and same amount of memory}
\begin{figure*}
    \centering
    \includegraphics[width = 0.65\textwidth]{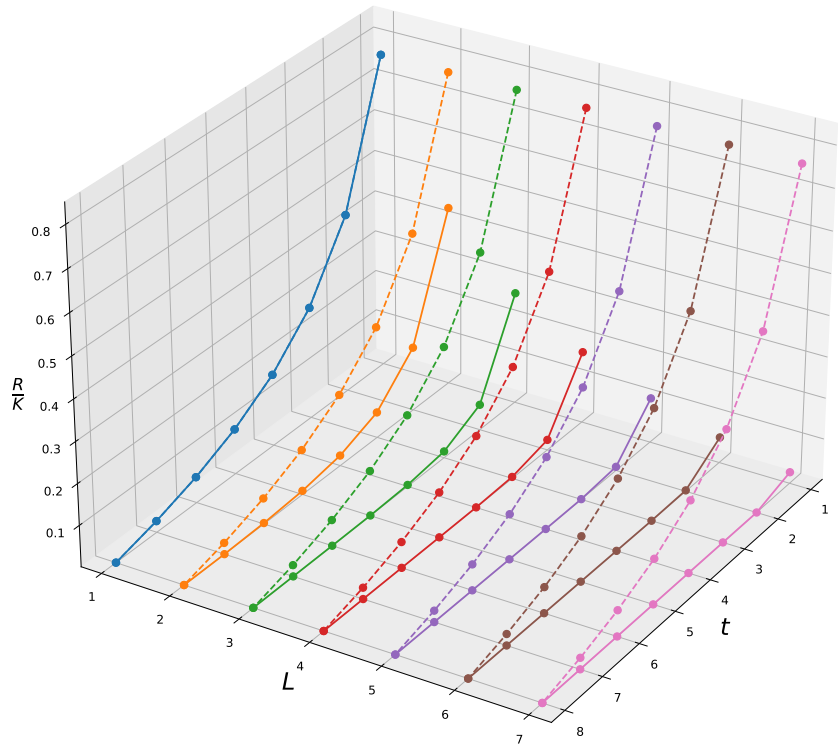}
    \caption{$\frac{R}{K}$ for Dedicated cache (dotted lines) and Multi Access (solid lines). Here $C_{MA} = C_{DC} = 8$ and $t_{MA} = t_{DC} = t$}%
    \label{C=K}
\end{figure*}

\begin{figure*}
    \centering
    \includegraphics[width = 0.65\textwidth]{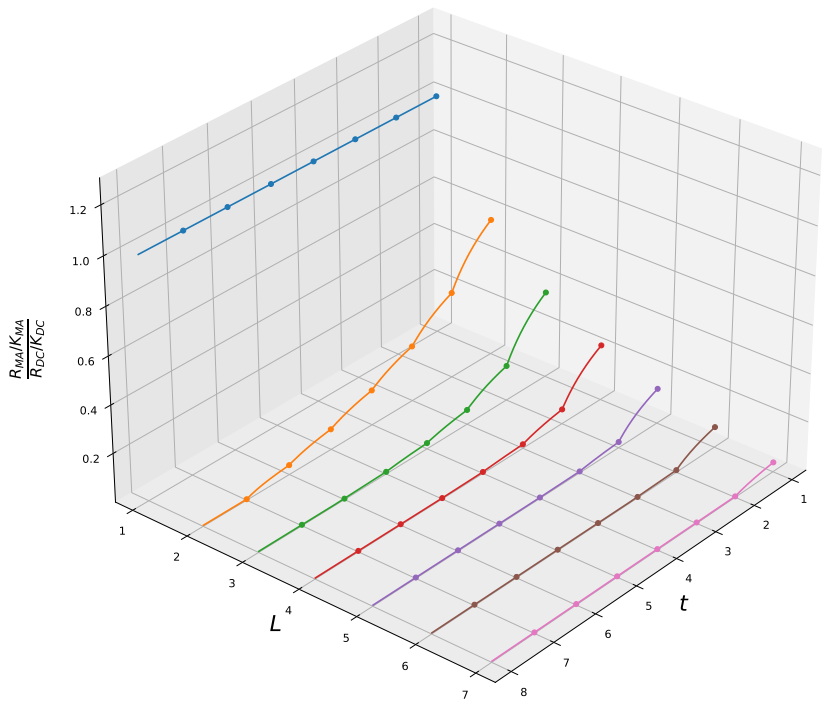} 
    \caption{Ratio of per user rate $\frac{R_{MA} / K_{MA}}{R_{DC} / K_{DC}}$. Here $C_{MA} = C_{DC} = 8$ and $t_{MA} = t_{DC} = t$}%
    \label{C=K2}
\end{figure*}

        As the number of cache nodes in the multi-access system is $C_{MA}$ and in the dedicated cache system, the number of cache nodes are same as the number of users $K_{DC}$, we will consider $C_{MA} = K_{DC} = 8$ i.e.\ $8$ cache nodes in both, dedicated cache and multi-access setup. We will also consider that the total memory of both systems are also same i.e.\ $t_{MA} = t_{DC} = t$. Let the cache access degree for the multi-access system be  $L$. Note that $K_{MA} = \binom{C}{L}$, i.e.\ the number of users in the multi-access setup will be higher than the dedicated cache system except for $L = 1$ where both systems are same. So with same amount of memory, multi- access setup is able to support higher number of users. We will compare rate per users for the two setups. Note that although $t_{MA} = t_{DC}$, users in multi-access setup are accessing $L$ cache nodes, therefore they have access to more cache memory than the users of the dedicated cache setting. The rate per user for both systems are plotted in Figure~\ref{C=K} for $8$ cache nodes in each setting for different values of $t$ and $L$. We can see that per user rate of the multi-access setup is better than that of the dedicated cache setup. Ratio of per user rates is plotted in Figure~\ref{C=K2}. Due to large number of users in multi-access setup, a single transmission from a server can benefit more number of users than the number of users benefited by a single transmission in dedicated cache setup. We will see that in the multi-access system a single transmission is simultaneously used by $\binom{L+t}{L}$ users, compared to $t+1$ in the corresponding dedicated cache setup. 

    \subsubsection{Same number of cache and same memory per user}
\begin{figure*}
    \centering
    \includegraphics[width = 0.65\textwidth]{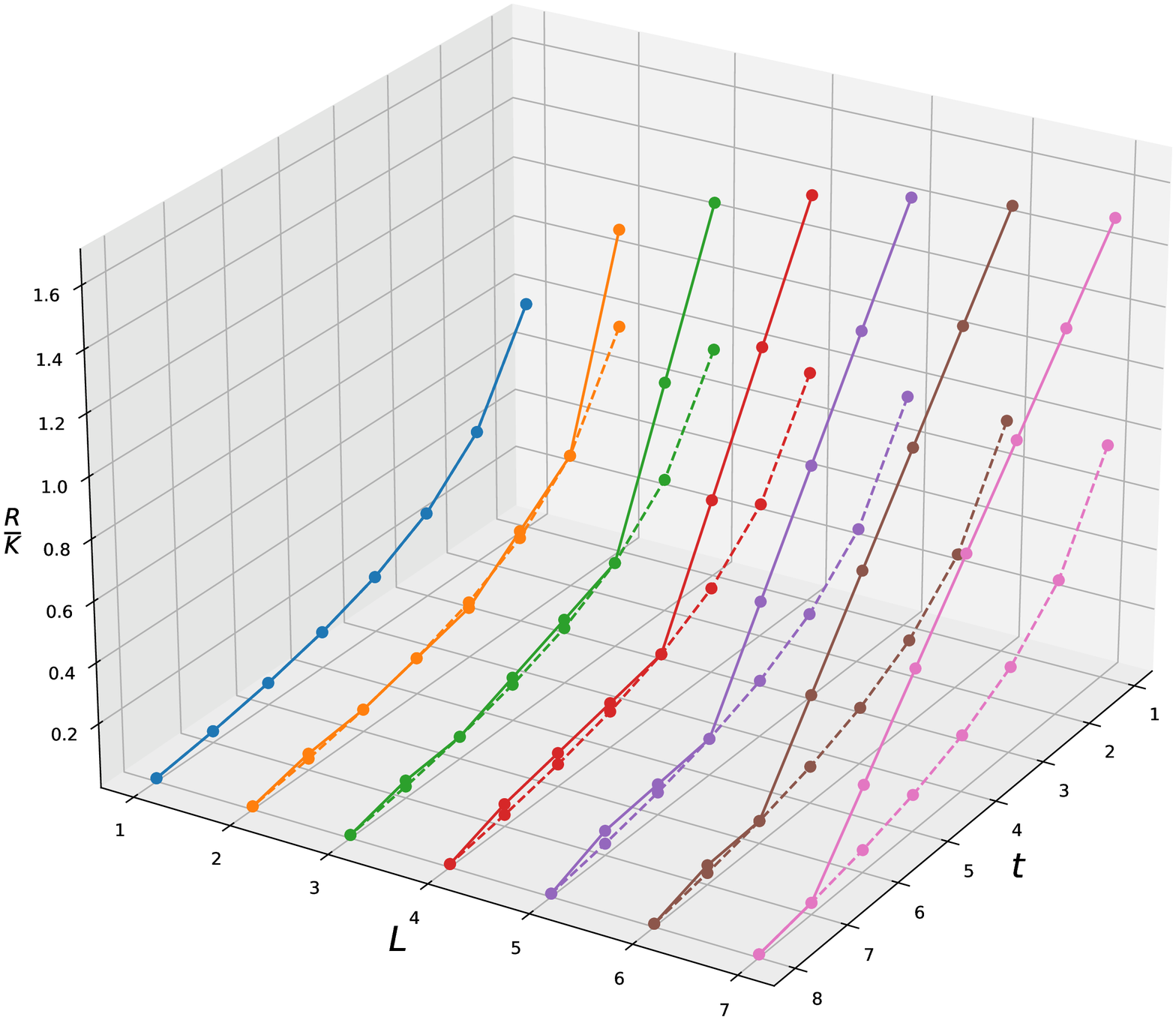}
    \caption{$\frac{R}{K}$ for Dedicated cache (dotted lines) and Multi Access (solid lines). Here $C_{MA} = C_{DC} = 8$ and $t_{DC} = Lt_{MA} = t$}%
    \label{C=K_tneqt}
\end{figure*}

\begin{figure*}
    \centering
    \includegraphics[width = 0.65\textwidth]{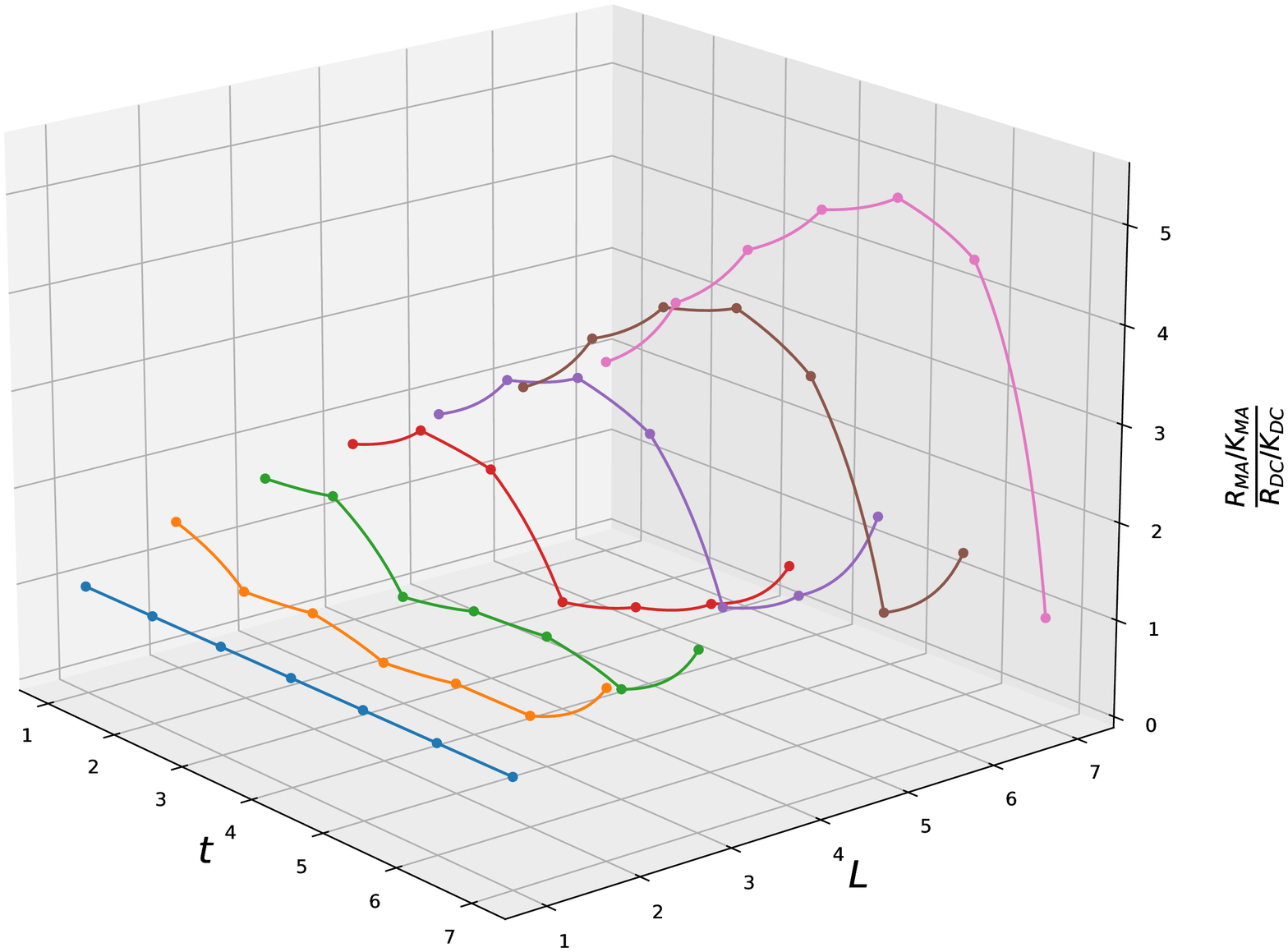} 
    \caption{Ratio of per user rate $\frac{R_{MA} / K_{MA}}{R_{DC} / K_{DC}}$. Here $C_{MA} = C_{DC} = 8$ and $t_{DC} = Lt_{MA} = t$}%
    \label{C=K_tneqt2}
\end{figure*}

        Users in multi-access setup are accessing more memory than users of dedicated cache setting if $C_{MA} = K_{DC}$ and $t_{MA} = t_{DC}$. Now, we reduce the storage capacity in multi-access setup so that amount of memory accessed by users of dedicated cache setup and multi-access setup is same. We consider that $C_{MA} = K_{DC} = 8$ but this time storage capacity of cache nodes in the multi-access setup is smaller than that of the dedicated cache system, so that each user is accessing same amount of memory. For that we set $M_{DC} = L \times M_{MA}$ because every user in the multi-access setup is accessing $L$ cache nodes. Therefore we have 
        \[
            \frac{t_{DC}N}{K_{DC}} = L\frac{t_{MA}N}{C_{MA}} \implies t_{MA} = t_{DC} / L.
        \]
        Now, although each user has access to the same amount of memory, each user of a dedicated cache system has access to a cache whose content is independent of the content stored in cache of other users. Since the cache content accessed by the users of the multi-access setup is not independent of the cache content accessed by other users, therefore the rate of the dedicated cache setup will be better than the rate of the multi-access setup. In Figure~\ref{C=K_tneqt} we compare per user rate for $L$ and $t_{DC}$ ranging in $[C]$. Although, even for per user rate we see that the dedicated system is performing better than the multi-access setup for most of the cases, we see that for cases when $t_{DC} = L$ per user rate of both settings coincide as
        \[
            \frac{R_{MA} / K_{MA}}{R_{DC} / K_{DC}} = \frac{ \frac{1}{\binom{C}{L}} \frac{\binom{C}{L+1}}{C} }{ \frac{1}{C} \frac{\binom{C}{t_{DC}+1}}{\binom{C}{t_{DC}}} } = 1.
        \]
        Beyond that, there also exist some points where, the per user rate of the multi-access setup is better than that of the dedicated cache setup, for instance in Figure~\ref{C=K_tneqt} when $t_{DC} = 4$ and $L = 2$ we have $\frac{R_{DC}}{K_{DC}} = \frac{1}{10}$ whereas $\frac{R_{MA}}{K_{MA}} = \frac{1}{11.2}$. Ratio of per user rates is plotted in Figure~\ref{C=K_tneqt2}

\begin{figure*}
    \centering
    \includegraphics[width = 0.6\textwidth]{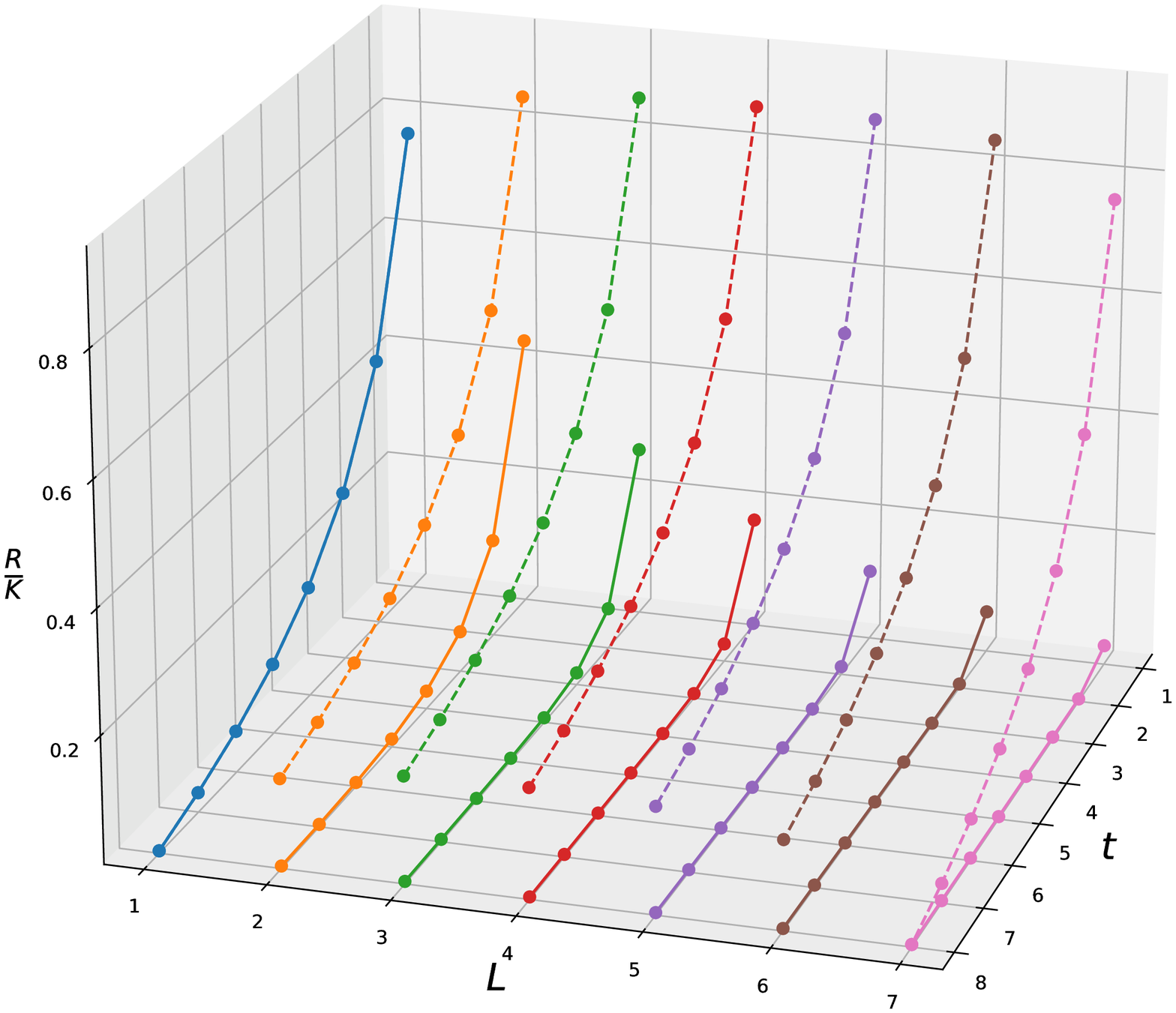}
    \caption{$\frac{R}{K}$ for Dedicated cache (dotted lines) and Multi Access (solid lines). Here $C_{MA} = 8$, $K_{MA} = K_{DC} = \binom{C_{MA}}{L}$ and $t_{DC} = t_{MA} = t$}%
    \label{K=K}
\end{figure*}

\begin{figure*}
    \centering
    \includegraphics[width = 0.65\textwidth]{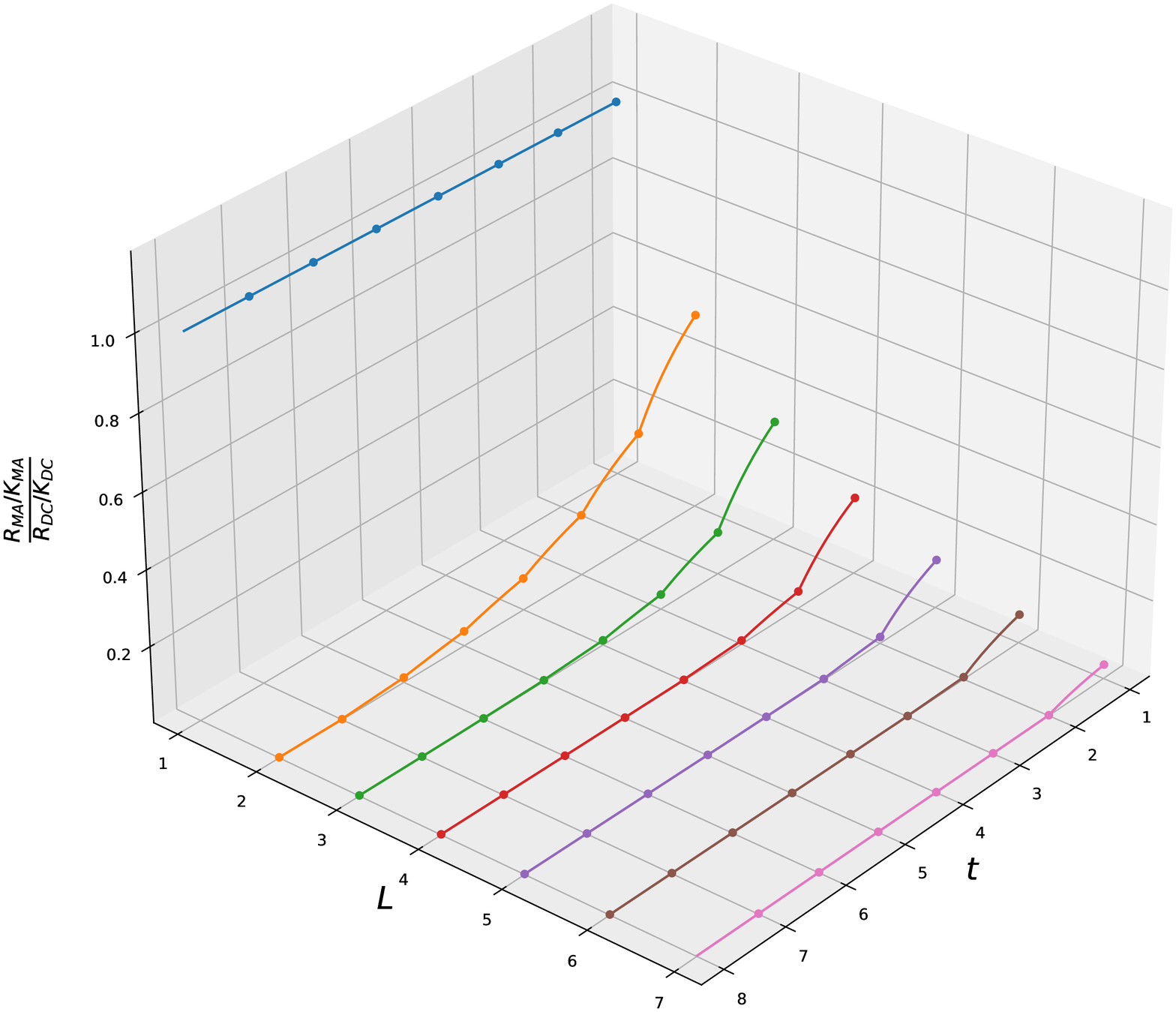} 
    \caption{Ratio of per user rate $\frac{R_{MA} / K_{MA}}{R_{DC} / K_{DC}}$. Here $C_{MA} = C_{DC} = 8$ and $t_{DC} = t_{MA} = t$}%
    \label{K=K2}
\end{figure*}

\begin{figure*}
    \centering
    \includegraphics[width = 0.65\textwidth]{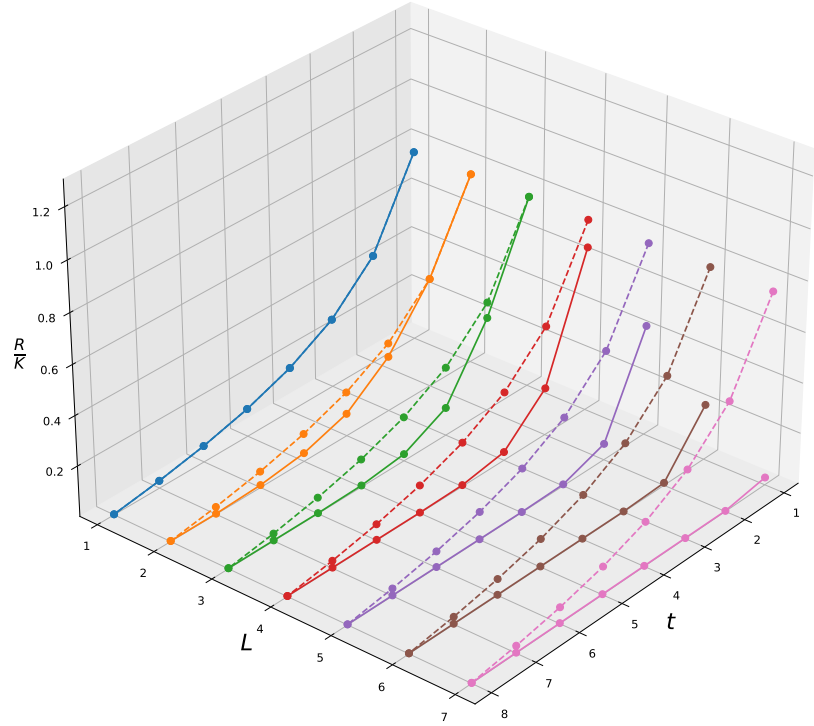}
    \caption{$\frac{R}{K}$ for Dedicated cache (dotted lines) and Multi Access (solid lines) with cyclic wraparound cache access. Here $C_{MA} = 8$, $K_{MA} = K_{DC} = 8$ and $t_{DC} = t_{MA} = t$}%
    \label{cyclic}
\end{figure*}

\begin{figure*}
    \centering
    \includegraphics[width = 0.7\textwidth]{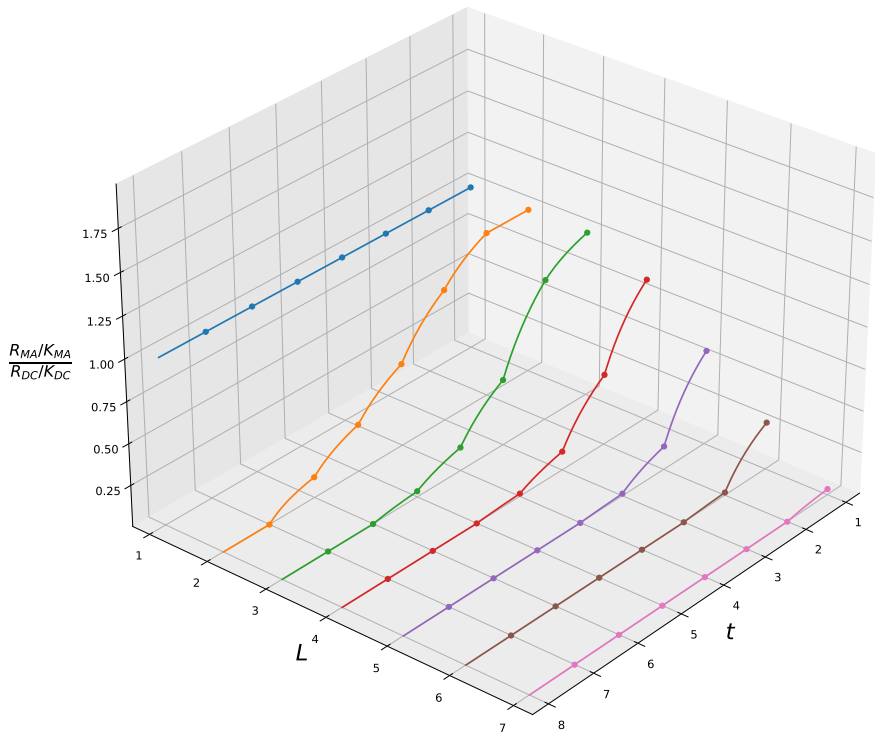} 
    \caption{Ratio of per user rate $\frac{R_{MA} / K_{MA}}{R_{DC} / K_{DC}}$.  Here $C_{MA} = 8$, $K_{MA} = K_{DC} = 8$ and $t_{DC} = t_{MA} = t$ and users are accessing cache in cyclic wraparound fashion in multi-access setup.}%
    \label{cyclic2}
\end{figure*}

    \begin{table*}[tbh]
        \centering
        \begin{minipage}{0.4\textwidth}\label{Qtable}
            \caption{Query Table}
            \begin{eqnarray*}
                \begin{array}{|c|c|c|c|c|}\hline
                \mbox{Server 1}&\mbox{Server 2}\\\hline
                a_1,b_1,c_1&a_2,b_2,c_2\\\hline
                a_3,b_2&a_4,b_1\\
                b_3,c_3&b_4,c_4\\
                a_5,c_2&a_6,c_1\\\hline
                a_7 , b_4,c_4&a_8 , b_3,c_3\\
                \hline
                \end{array}
            \end{eqnarray*}
        \end{minipage}
        \begin{minipage}{0.55\textwidth}\label{Atable}
            \caption{Answer from servers}
            \begin{eqnarray*}
                \begin{array}{|c|c|c|c|c|}\hline
                \mbox{ Server 1}&\mbox{ Server 2}\\\hline
                    W_{1 , \{4 , 5\}}^{a_1},W_{2 , \{4 , 5\}}^{b_1},W_{3 , \{4 , 5\}}^{c_1}&W_{1 , \{4 , 5\}}^{a_2},W_{2 , \{4 , 5\}}^{b_2},W_{3 , \{4 , 5\}}^{c_2}\\\hline
                    W_{1 , \{4 , 5\}}^{a_3}+W_{2 , \{4 , 5\}}^{b_2}&W_{1 , \{4 , 5\}}^{a_4}+W_{2 , \{4 , 5\}}^{b_1}\\
                    W_{2 , \{4 , 5\}}^{b_3}+W_{3 , \{4 , 5\}}^{c_3}&W_{2 , \{4 , 5\}}^{b_4}+W_{3 , \{4 , 5\}}^{c_4}\\
                    W_{1 , \{4 , 5\}}^{a_5}+W_{3 , \{4 , 5\}}^{c_2}&W_{1 , \{4 , 5\}}^{a_6}+W_{3 , \{4 , 5\}}^{c_1}\\\hline
                    W_{1 , \{4 , 5\}}^{a_7}+W_{2 , \{4 , 5\}}^{b_4}+W_{3 , \{4 , 5\}}^{c_4}&W_{1 , \{4 , 5\}}^{a_8}+W_{2 , \{4 , 5\}}^{b_3}+W_{3 , \{4 , 5\}}^{c_3}\\
                \hline
                \end{array}
            \end{eqnarray*}
        \end{minipage}
    \end{table*}

    \subsubsection{Same number of users and same total system memory}
        Keeping the number of users in both the setups same we have $K_{DC} = K_{MA} = \binom{C_{MA}}{L}$. Also we will keep the total memory in both systems same i.e.\ $t_{MA} = t_{DC} = t$. Note that the number of cache nodes in the  dedicated cache system is $K_{DC} = \binom{C_{MA}}{L}$. For the same total memory, we want that 
        \[
            C_{MA}M_{MA} = K_{DC}M_{DC} \implies M_{DC} = \frac{C_{MA}}{\binom{C_{MA}}{L}} M_{MA}. 
        \]
        So, the size of an individual cache node in the dedicated cache system is now smaller than that of the multi-access system. Then again, users of the multi-access setup can access more than one cache node. Therefore, we see that the dedicated cache system have two disadvantages. For supporting the same number of users as the multi-access setup, using same total system memory, the size of individual cache have to be reduced and every user get a smaller share of total memory compared to multi-access system. The comparison of per user rate for this setting is shown in Figure~\ref{K=K}. Note that because the number of user in both the settings are kept equal the ratio of per user rate will also be same as ratio of rate in both the settings. Ratio of per user rates is plotted in Figure~\ref{K=K2}. We can see that the multi-access system is performing better than the dedicated cache setup while utilising the same amount of cache memory, and serving the same number of users.

    \subsubsection{Multi Access setup with cyclic wraparound cache access}
        Finally, we consider multi-access setup with cyclic wraparound access. In this setting $C_{MA} = K_{MA} = C$ and user $k$ will access cache nodes $\{ k , k+1 , \cdots , k+L-1 \}$ where sum is modulo $C$ except $k+l = C$ if $k+l$ is multiple of $C$. We compare this to the dedicated cache setup with $K_{DC} = C$ users. We also consider cache sizes to be identical in both the settings i.e.\ $M_{MA} = M_{DC} = M$. Along with the same number of cache nodes, this will imply that $t_{DC} = t_{MA}$. Users in multi-access setup are accessing the caches which users of dedicated cache setup are also accessing. So, the rate of the multi-access setup will be no higher than the rate of the dedicated cache setup. For instance, for given memory points $M_1$ and $M_2$, if the rate of the multi-access setup is $R_{MA1}$ and $R_{MA2}$ respectively and the rate of the dedicated cache setup is $R_{DC1}$ and $R_{DC2}$ respectively with $R_{MA1} > R_{DC1}$ and $R_{MA2} < R_{DC2}$ then, in multi-access system, better rate can be achieved by using storage and coding strategy of dedicated cache system at memory point $M_1$ and by using memory sharing between $M_1$ and $M_2$. We compare per user rate of both the settings in Figure~\ref{cyclic}. Here we see that due to access to more cache memory multi-access system is performing better than the dedicated cache setup for large $L$. Due to the same number of users in both the settings, ratio of per user rate will be same as ratio of rates in two settings. In Figure~\ref{cyclic2} we plot ratio of per user rate in both the settings. 

\section{Achievable Scheme in Theorem \ref{Point_rates}}
\label{scheme}
In the following subsection we discuss an example which will help in understanding the general description in the following  subsection.  
\subsection{Example}

    Consider the multi-access setup with $S = 2$ non-colluding servers, $N = 3$ files $W_1 , W_2$ and $W_3$. There are $C = 5$ cache nodes, each node capable of storing $M / N = 2 / 5$ fraction of each file. There are $K = 10$ users and each user is connected to a unique set of $L = 3$ cache nodes. We will index the users with the indices of cache nodes they are connected to, for example user $\{2 , 4\}$ is connected to cache node $2$ and cache node $4$. The caches will be filled in placement phase as follows

    \textbf{Placement Phase}: Let $t = \frac{CM}{N} = 2$. Divide each file into $\binom{C}{t} = \binom{5}{2} = 10$ subfiles.
    \[
        W_n = \left\{W_{n , \mathcal{T}} | \mathcal{T} \in \binom{[5]}{2}\right\}
    \]

    and fill cache node $c \in [5]$ as follows
    \[
        Z_c = \{W_{n , \mathcal{T}} | n \in [N] , \forall \mathcal{T} \in \binom{[5]}{2} \mbox{ such that } c \in \mathcal{T} \}.
    \]
    
    \textbf{Delivery Phase}: Now every user will choose a file index, and want to retrieve the file from the servers privately. Let the demand of user $\mathcal{K} \in \binom{[5]}{2}$ be $d_{\mathcal{K}}$. Then the demand vector is $\textbf{d} = {(d_{\mathcal{K}})}_{\mathcal{K} \in \binom{[5]}{3}}$, and users want to hide this demand vector from the servers. For this, each subfile will be further divided into $S^N = 8$ sub-subfiles and users will generate $2$ queries $\textbf{Q}_1^{\textbf{d}}$ and $\textbf{Q}_2^{\textbf{d}}$ one for each server as follows. For every $\mathcal{S} \in \binom{[C]}{t+L} = \binom{[5]}{5} = \{\{1 , 2 , 3 , 4 , 5\}\}$ generate
    \[
        \textbf{Q}_s^{\textbf{d} , \mathcal{S}} = \left\{ Q_s^{d_{\mathcal{K}} , \mathcal{S}} | \mathcal{K} \in \binom{\mathcal{S}}{3} \right\}
    \]
    where $Q_s^{d_{\mathcal{K}} , \mathcal{S}}$ is the query sent to server $s$ in single user PIR setup, with $S$ servers and $N$ files if the demand of the user is $d_{\mathcal{K}}$. For instance, consider $\mathcal{K} = \{1 , 2 , 3\}$ and $d_{\{1 , 2 , 3\}} = 1$ then in order to generate $Q_1^{1 , \mathcal{S}} , Q_2^{1 , \mathcal{S}}$ three random permutations of $[S^N] = 8$ will be formed, one corresponding to each file. Let these permutations be $\{a_1 \hdots a_8\} , \{b_1 \hdots b_8\}$ and  $\{c_1 \hdots c_8\}$ for files $W_1 , W_2$ and $W_3$ respectively. Now, as user $\{1 , 2 , 3\}$ wants the file $W_1$ then the query $Q_s^{1 , \mathcal{S}}$ will be a list of sub-subfile index of subfiles $\{4 , 5\}$ as given in Table I. %~\ref{Qtable}
    These lists will be generated for every $\mathcal{K} \in \binom{[5]}{3}$ and queries will be sent to the respective server. After receiving the query server $s$ will transmit 
    \[
    \bigoplus_{\mathcal{K} \in \binom{S}{3}} A_s^{d_{\mathcal{K}}}( Q_s^{d_{\mathcal{K}} , \mathcal{S}} , W_{[3] , \mathcal{S} \setminus \mathcal{K}} )
    \]

    \noindent where $A_s^{d_{\mathcal{K}}}( Q_s^{d_{\mathcal{K}} , \mathcal{S}} , W_{[3] , \mathcal{S} \setminus \mathcal{K}} )$ is the answer of server $s$ in single user PIR setup if query is $Q_s^{d_{\mathcal{K}} , \mathcal{S}}$ and the set of files is $W_{[3] , \mathcal{S} \setminus \mathcal{K}}$. Again, considering $\mathcal{K} = \{1 , 2 , 3\}$, $A_s^{1}( Q_s^{1 , \mathcal{S}} , W_{[3] , \mathcal{S} \setminus \{1 , 2 , 3\}} )$ is given in Table II. %~\ref{Atable}
    After listening to the broadcast from the servers, every user will be able to decode their desired subfile. Again considering the case of user $\{1 , 2 , 3\}$, it has access to all subfiles of all files indexed by $\{1,2\} , \{1,3\} , \{1,4\} , \{1,5\} , \{2,3\} , \{2,4\} , \{2,5\} \mbox{ and } \{3,4\}$. Only missing subfiles are those indexed by $\{4,5\}$. Now consider again the transmission of the server
    \begin{align*}
        &\bigoplus_{\mathcal{K} \in \binom{S}{3}} A_s^{d_{\mathcal{K}}}( Q_s^{d_{\mathcal{K}}} , W_{[3] , \mathcal{S} \setminus \mathcal{K}} ) \\ 
        =& A_s^{1}( Q_s^{1} , W_{[3] , \{4 , 5\}} ) + \bigoplus_{\mathcal{K} \in \binom{S}{3} \setminus \{1 , 2 , 3\}} A_s^{d_{\mathcal{K}}}( Q_s^{d_{\mathcal{K}}} , W_{[3] , \mathcal{S} \setminus \mathcal{K}} ).
    \end{align*}
    User $\{1 , 2 , 3\}$ has access to all the subfiles indexed by $\mathcal{S} \setminus \mathcal{K}, \forall \mathcal{K} \in \binom{S}{3} \setminus \{1 , 2 , 3\}$, so it can cancel out the summation term above and will be left with $A_1^{1}( Q_s^{1 , \mathcal{S}} , W_{[3] , \{4 , 5\}} )$ and $A_2^{1}( Q_s^{1 , \mathcal{S}} , W_{[3] , \{4 , 5\}} )$. The user will get all sub-subfiles of $W_{1 , \{4 , 5\}}$ from these remaining terms. Similarly all the users will get the missing subfiles of the demanded file. Also, to generate ${\{ Q_s^{d_{\mathcal{K}} , \mathcal{S}}\}}_{s \in [S]}$ independent random permutations of $[8]$ is chosen for every $\mathcal{S}$. Then, from the privacy of single user PIR scheme, servers will get no information about the demand vectors from the queries they got.

    Also note that, as each server is transmitting $7$ sub-subfiles each of size $\frac{1}{10 \times 8}$ units, the rate in this example is $R = \frac{7}{40}$ and subpacketization level is $80$

    \subsection{General Description}\label{general_scheme}

   Consider $N$ unit size files ${\{W_n\}}_{n \in [N]}$ replicated across $S$ servers. There are $C$ cache nodes each capable of storing $M$ files and $K$ users each connected to a unique set of $L$ cache nodes. We will consider the system with $\binom{C}{L}$ users. As each user is connected to a unique set of $L$ cache, we will index that user with a $L$ sized subset of $[C]$. Specifically, user $\mathcal{K}$, where $\mathcal{K} \in \binom{[C]}{L}$, is the user connected to cache nodes indexed by $\mathcal{K}$.

   \textbf{Placement Phase}: Let $t = \frac{C M}{N}$ be an integer. Then divide each file into $\binom{C}{t}$ subfiles each indexed by a $t$ sized subset of $[C]$ as
    \[
        W_n = \left\{W_{n , \mathcal{T}} | \mathcal{T} \in \binom{[C]}{t} \right\}.
    \]

    Then fill cache node $c$ with
    \[
        Z_c = \left\{W_{n , \mathcal{T}} | c \in \mathcal{T} , \mathcal{T} \in \binom{[C]}{t}\right\}.
    \]
    \textbf{Delivery Phase}: In this phase every user will choose one of the file index. Let user $\mathcal{K} , \forall \mathcal{K} \in \binom{[C]}{L}$ choose index $d_{\mathcal{K}} \in [N]$. User ${\mathcal{K}}$ will then wish to retrieve file $W_{d_{\mathcal{K}}}$ from the servers without revealing the index of the demanded file to the servers. Let, $\textbf{d} = {(d_{\mathcal{K}})}_{\mathcal{K} \in \binom{[C]}{L}}$ be the demand vector. Users don't want the servers to get any information about the demand vector. For privately retrieving the files, the users will cooperatively generate $S$ queries $\textbf{Q}_s^{\textbf{d}}$ as follows. For every $\mathcal{S} \in \binom{[C]}{t+L}$, users will generate 
    \[
        \textbf{Q}_{s}^{\textbf{d} , \mathcal{S}} =  \left\{Q_s^{d_{\mathcal{K}} , \mathcal{S}} | \mathcal{K} \in \binom{\mathcal{S}}{L}\right\} 
    \]
    \noindent and the query sent to server $s$ will be
    \[
        \textbf{Q}_s^{\textbf{d}} = {\{\textbf{Q}_{s}^{\textbf{d} , \mathcal{S}}\}}_{\mathcal{S} \in \binom{[C]}{t+L}} 
    \]
    where $Q_s^{d_{\mathcal{K}} , \mathcal{S}}$ is the query sent to server $s$ in the single user PIR setup of~\cite{Sun17PIR} if the user demand is $d_{\mathcal{K}}$.
 
    Now for every $\textbf{Q}_{s}^{\textbf{d} , \mathcal{S}}$ server $s$ will transmit
    \[
        \bigoplus_{\mathcal{K} \in \binom{\mathcal{S}}{L}} A_s^{d_{\mathcal{K}}}(Q_s^{d_{\mathcal{K}} , \mathcal{S}} , W_{[N] , \mathcal{S} \setminus \mathcal{K}})
    \]
    where $A_s^{d_{\mathcal{K}}}(Q_s^{d_{\mathcal{K}} , \mathcal{S}} , W_{[N] , \mathcal{S} \setminus \mathcal{K}})$ is the answer of server $s$ in single user PIR setup if received query is $Q_s^{d_{\mathcal{K}}}$ and set of files is $\{W_{[N] , \mathcal{S} \setminus \mathcal{K}}\}$.

    Now we will see that all the users will be able to decode their required file from these transmissions and the caches they have access to.
    
\subsubsection*{Decoding}
    Consider user $\mathcal{K}$ (i.e.\ the user connected to cache nodes indexed by $\mathcal{K}$) and subfile index $\mathcal{T}$. If $\mathcal{K} \cap \mathcal{T} \neq \phi$ then subfile $W_{d_{\mathcal{K}} , \mathcal{T}}$ is available to the user from the cache. If $\mathcal{K} \cap \mathcal{T} = \phi$ then the subfile has to be decoded from the transmissions. Consider the transmissions corresponding to $\mathcal{S} = \mathcal{K} \cup \mathcal{T}$.
  \begin{align*}
        &\bigoplus_{\mathcal{K}^{\prime} \in \binom{\mathcal{K} \cup \mathcal{T}}{L}} A_s^{d_{\mathcal{K}^{\prime}}}(Q_s^{d_{\mathcal{K}} , \mathcal{K} \cup \mathcal{T}} , W_{[N] , (\mathcal{K} \cup \mathcal{T}) \setminus \mathcal{K}^{\prime}}) \\
        &= A_s^{d_{\mathcal{K}}}(Q_s^{d_{\mathcal{K}} , \mathcal{K} \cup \mathcal{T}} , W_{[N] , \mathcal{T}}) \oplus \\
        &\bigoplus_{\mathcal{K}^{\prime} \in \binom{\mathcal{K} \cup \mathcal{T}}{L} \setminus \mathcal{K}} A_s^{d_{\mathcal{K}^{\prime}}}(Q_s^{d_{\mathcal{K}^{\prime}} , \mathcal{K} \cup \mathcal{T}} , W_{[N] , (\mathcal{K} \cup \mathcal{T}) \setminus \mathcal{K}^{\prime}})
  \end{align*}
   User $\mathcal{K}$ has access to all the subfiles in the second term of RHS above, and therefore it can recover the first term from the above expression. After getting $A_s^{d_{\mathcal{K}}}(Q_s^{d_{\mathcal{K}} , \mathcal{K} \cup \mathcal{T}} , W_{[N] , \mathcal{T}})$ for all $s \in [S]$, user $\mathcal{K}$ can recover subfile $W_{d_{\mathcal{K}} , \mathcal{T}}$ from the transmissions.

\subsubsection*{Rate}
    Each server is performing $\binom{C}{t+L}$ transmissions each of size $\frac{1}{\binom{C}{t}} \big( \frac{1}{S} + \frac{1}{S^2} + \cdots + \frac{1}{S^{N}} \big)$ units. So the rate is
  \[
      R(t) = \frac{ \binom{C}{t+L} }{ \binom{C}{t} } \left( 1 + \frac{1}{S} + \frac{1}{S^2} + \hdots + \frac{1}{S^{N-1}} \right).
  \]
\subsubsection*{Subpacketization} 
    As we can see each file is divided into $\binom{C}{t}$ subfiles, each of which have to be further divided into $S^N$ sub-subfiles. So the subpacketization level is $\binom{C}{t} \times S^N.$

\subsubsection*{Coding Gain}
    We can see that transmission corresponding to each $\mathcal{S} \in \binom{[C]}{t+L}$ is beneficial to user $\mathcal{K}$ if $\mathcal{K} \in \binom{\mathcal{S}}{L}$. So every transmission, from each server, is used by $\binom{L+t}{L}$ users. 

\subsubsection*{Proof of Privacy}\label{privacy_proof}
    \begin{figure*}[hbt!]
        \centering
        \includegraphics[width = 0.9\textwidth]{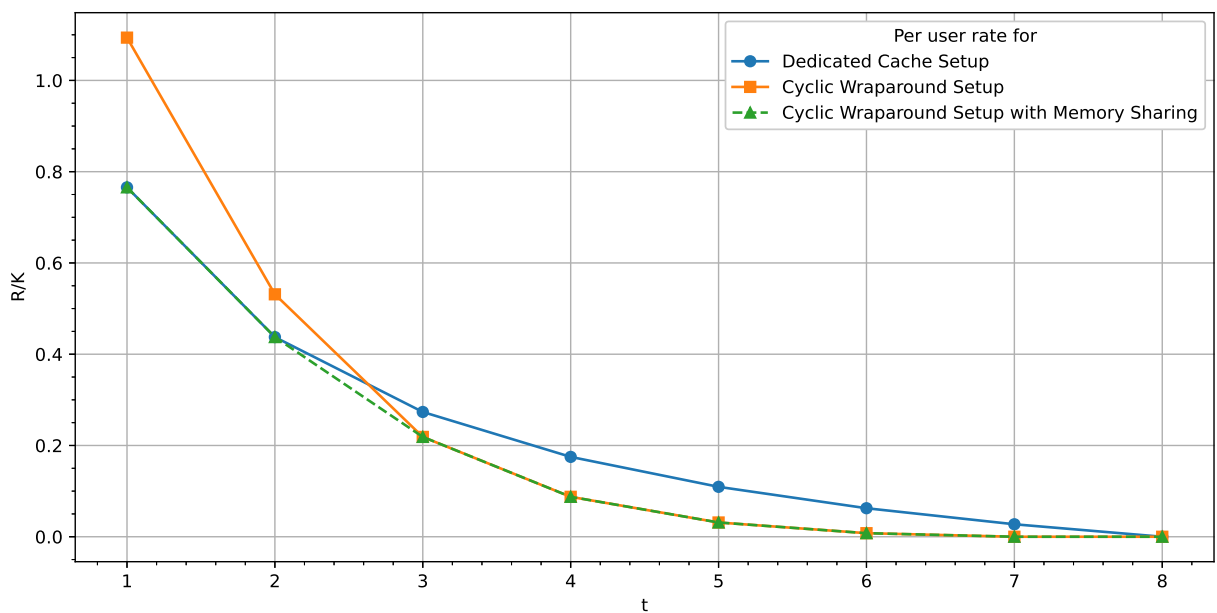}
        \caption{Per user rate for $C = 8 , L = 2 , S = 2 , N = 3$. Multi Access setup with cyclic wraparound cache access achieve rate only as high as dedicated cache setup with equal total memory in both systems.}%
        \label{cyc_memory_sharing}
    \end{figure*}
    Now we will prove that none of the server will get any information about the demand vector. We will show that given any realisation of query sent to server $s$ say $\textbf{Q}_s^{\textbf{d}} = \textbf{q}_s$ all possible demand vectors are equally likely. Consider
    \[
        \mathbb{P}(\textbf{d} = d_{[K]} | \textbf{Q}_s^{\textbf{d}} = \textbf{q}_s) = \frac{\mathbb{P}(\textbf{Q}_s^{\textbf{d}} = \textbf{q}_s | \textbf{d} = d_{[K]})\mathbb{P}(\textbf{d} = d_{[K]})}{\mathbb{P}(\textbf{Q}_s^{\textbf{d}} = \textbf{q}_s)}
    \]
    for some $d_{[K]} \in {[N]}^K$. Query $\textbf{q}_s$ sent to server $s$ consist of
    \[
        \textbf{q}_s = \left\{ \textbf{q}_s^{\mathcal{S}} | \mathcal{S} \in \binom{[C]}{t+L} \right\}
    \]
    and each $\textbf{q}_s^{\mathcal{S}}$ consist of $\binom{t+L}{L}$ lists of sub-subfile indices. Each list of indices is made using an independent random permutation of $[S^N]$. In each list, there are $S^{N-1}$ unique sub-subfile indices corresponding to each file. Now, the indices corresponding to the user demand never repeat across the servers and a given server only sees $S^{N-1}$ unique indices corresponding to each file in every list. There are ${(S^N - S^{N-1})!}^N$ permutations that'll result in the same list of indices. As all $\binom{t+L}{L}$ lists in $\textbf{q}_s^{\mathcal{S}}$ are generated independently and as for every $\mathcal{S} \in \binom{[C]}{L + t}$ the choice of $\textbf{q}_s^{\mathcal{S}}$ are independent, we have 
    \[
        \mathbb{P}(\textbf{Q}_s^{\textbf{d}} = \textbf{q}_s | \textbf{d} = d_{[K]}) = {\frac{(S^N - S^{N-1})!}{S^N !}}^{N \binom{t+L}{L} \binom{C}{t+L}}
    \]
    which does not depend on the demand vector. Therefore
    \[
        \mathbb{P}(\textbf{d} = d_{[K]} | \textbf{Q}_s^{\textbf{d}} = \textbf{q}_s) = \frac{1}{N^K} = \mathbb{P}(\textbf{d} = d_{[K]})
    \]
    which means every possible demand vector is equiprobable given the query to a server. Hence, servers will not get any information about demand vector.

\section{Cyclic wraparound cache access}\label{cyclic_scheme}
    The achievable scheme of Section~\ref{scheme} is for the case when there are $\binom{C}{L}$ users and every user is accessing a unique set of $L$ cache. But the same transmission strategy will also work if number of users are less than $\binom{C}{L}$, provided every user is accessing a unique set of $L$ cache nodes. Furthermore, it is possible to achieve correctness and privacy with reduced transmissions. Let there are $K \leq \binom{C}{L}$ users and consider that the user $k$ is connected to caches indexed by $\mathcal{K}_k$, where $\mathcal{K}_k \in \binom{[C]}{L}$ and $\mathcal{K}_k \neq \mathcal{K}_{k^{\prime}}$ for $k \neq k^{\prime}$ for every $k , k^{\prime} \in [K]$. Now, user $k$ will get its required subfiles from the transmissions corresponding to subsets $\mathcal{S} \in \binom{[C]}{L+t}$ if $\mathcal{K}_k \subset \mathcal{S}$. Transmission corresponding to subset $\mathcal{S}$ will be beneficial to user $k$ only if $\mathcal{K}_k \subset \mathcal{S}$. So, if there exist some $\mathcal{S} \in \binom{[C]}{t+L}$ such that $\mathcal{K}_k \not\subset \mathcal{S}$ for all $k \in [K]$ then transmission corresponding to $\mathcal{S}$ can be avoided. Therefore, only transmissions required are the ones corresponding to those $\mathcal{S} \in \binom{[C]}{t+L}$ for which there is atleast one $k \in [K]$ such that $\mathcal{K}_k \subset \mathcal{S}$.

    Now consider cyclic wraparound access with $C$ users and $C$ cache nodes. User $k$ is accessing cache nodes indexed by $\mathcal{K}_k = \{ k , k+1 , \cdots , k+L-1 \}$ where addition is modulo $C$ except $k+l = C$ if $k+l$ is multiple of $C$. In this case only those transmissions are performed that correspond to subsets 
    \[
        \Big\{\mathcal{S} \in \binom{[C]}{t+L} : \exists k \in [C] \mbox{ such that } \mathcal{S} \supset \mathcal{K}_k \Big\}.
    \]
    So only those $t+L$ sized subsets of $[C]$ are chosen which contain $L$ consecutive integers (with wrapping around $C$ allowed). This is same as the number of ways of choosing $t+L$ elements, from a set of $C$ distinguishable elements arranged in a circle, which contain atleast one subset of $L$ consecutive elements. As we will show in Section~\ref{cyc_proof}, there are $cyc(C , t+L , L)$ ways, as described in \eqref{cyceqn}, to choose $t+L$ caches out of $C$, such that there is atleast one subset of $L$ consecutive caches accessed by a user. Therefore in cyclic wraparound cache access setup only $cyc(C , t+L , L)$ transmissions are required for satisfying the user demands. In this case, user $k$ will get subfile $W_{d_k , \mathcal{T}}$ from the cache if $\mathcal{K}_k \cap \mathcal{T} \neq \phi$ and using the transmission corresponding to $\mathcal{K}_k \cup \mathcal{T}$ otherwise. 

    Also note that user $k$ of dedicated cache setup as well as multi-access setup with cyclic wraparound cache access is accessing cache node $k$. So, user $k$ of multi-access system can also decode the desired file from the transmissions for users of dedicated cache setup with same cache size, provided same cache contents in all caches. In dedicated cache setup $\binom{C}{t + 1}$ transmissions, each of size $\frac{1}{\binom{C}{t}}$ units, are required to satisfy the user demands. Therefore, when $cyc(C , t+L , L) > \binom{C}{t+1}$ we will perform placement and transmissions as done for dedicated cache setup. In this scenario the rate achieved in multi-access setup will only be as high as the rate achieved in dedicated cache scenario with same cache sizes. For $t \in [0:C]$ the rate achieved by multi-access setup would be     
    \[
        \min\Big\{ \frac{C - t}{t + 1} , \frac{cyc(C , t+L , L)}{\binom{C}{t}} \Big\} \left( 1 + \frac{1}{S} + \hdots + \frac{1}{S^{N-1}} \right).
    \]
    We will demonstrate this using the example for $C = 8$ and $L = 2$. In Figure~\ref{cyc_memory_sharing} we see that for smaller values of $t$ the cyclic wraparound cache access is incurring more per user rate than the dedicated cache setup. For instance, when $t = 2$, $cyc(8 , 4 , 2) = 68$ transmissions are performed for cyclic wraparound cache access without memory sharing (and incurring per user rate $0.531$) compared to $\binom{8}{3} = 56$ transmissions in dedicated cache setup (and incurring per user rate $0.437$). Therefore when $t = 2$, transmissions corresponding to dedicated cache setup will be performed. But when $t = 3$, multi-access system with cyclic wraparound cache access satisfy the user demand with $56$ transmissions (and incurring per user rate $0.219$) compared to dedicated cache setup which require $70$ transmissions (and incurring per user rate $0.273$), and therefore transmissions as described here will be performed. 
    \subsection{Proving the expression for $cyc(n , k , m)$}\label{cyc_proof}
        In this subsection we show that, number of ways of choosing $k$ integers from the set $[n]$, such that there is a subset of at least $m$ consecutive integers, with cyclic wrapping around $n$ allowed, is $cyc(n , k , m)$ as defined in \eqref{cyceqn}.
        %\begin{figure*}
        %\begin{align}
        %    cyc(n , k , m) &= \sum_{r=1}^k \Big(\binom{n-k}{r} + \binom{n-k-1}{r-1}\Big)\sum_{l=1}^r {(-1)}^{l-1}\binom{r}{l}\binom{k - l(m-1) -1}{r-1} \notag \\
        %    &+ \sum_{r=3}^k \binom{n-k-1}{r-2}\Big( \sum_{l=2}^{m-1}(l-1) \sum_{j = 1}^{r-2} {(-1)}^{j-1} \binom{r-2}{j} \binom{k - l - j(m-1) -1}{r-3} \notag \\
        %    &+ \sum_{l=m}^{k} (l-1) \binom{k-l-1}{r-3} \Big) + k-1 \label{Acyceqn}
        %\end{align}%
        %\end{figure*}
        
        First, for every $\Kappa \in \binom{[n]}{k}$, let $i_l$ denote the length of the $l^{th}$ consecutive runs of integers inside $\Kappa$ and $o_l$ denote the length of the $l^{th}$ consecutive run of integers outside $\Kappa$. For instance, if $n = 10$ and $\Kappa = \{1 , 2 , 4 , 9 , 10\}$ then $i_1 = 2$ corresponding to elements $\{1 , 2\}$ in $\Kappa$, $o_1 = 1$ corresponding to $\{3\}$ not in $\Kappa$, $i_2 = 1$ corresponding to element $\{4\}$ in $\Kappa$, $o_2 = 4$ corresponding to $\{5 , 6 , 7 , 8\}$ not in $\Kappa$ and $i_3 = 2$ corresponding to $\{9 , 10\}$ in $\Kappa$. Now every $\Kappa \in \binom{[n]}{k}$ can be uniquely determined by a sequence of positive integers consisting of $i_l$ and $o_l$, where every integer gives the length of consecutive runs of integers inside or outside $\Kappa$, provided it is known if $1$ is inside or outside $\Kappa$. For example, with $n=10$ and $k=6$ if we are given sequence of lengths of consecutive runs of integers inside and outside $\Kappa$ as $3 , 2 , 3 , 2$ and it is known that $1 \in \Kappa$ then, we can uniquely figure out $\Kappa = \{1 , 2 , 3 , 6 , 7 , 8\}$.

        Now, set of all $k$ sized subsets, $\Kappa$, of $[n]$ with at least one set of  $m$ cyclically consecutive integers, can be partitioned into four disjoint sets as follows
        \begin{enumerate}
            \item $1 \in \Kappa$ and $n \not \in \Kappa$. This correspond to sequences of the form $i_1 , o_1 , \hdots , i_r , o_r$ where $i_l , o_l \geq 1$ for all $l \in [r]$, $\sum_{l \in [r]} i_l = k$, $\sum_{l \in [r]} o_l = n-k$, $\exists l \in [r]$ such that $i_l \geq m$, $\forall r \in [k - m + 1]$. Let the set of all such $k$ sized subsets be denoted by $\Kappa_1$.
            \item $1 \not \in \Kappa$ and $n \in \Kappa$. This correspond to sequences of the form $o_1 , i_1 , \hdots , o_r , i_r$ where $i_l , o_l \geq 1$ for all $l \in [r]$, $\sum_{l \in [r]} i_l = k$, $\sum_{l \in [r]} o_l = n-k$, $\exists l \in [r]$ such that $i_l \geq m$, $\forall r \in [k - m + 1]$. Let the set of all such $k$ sized subsets be denoted by $\Kappa_2$.
            \item $1 \not \in \Kappa$ and $n \not \in \Kappa$. This correspond to sequences of the form $o_1 , i_1 , \hdots , o_r , i_r , o_{r+1}$ where $i_l , o_l \geq 1$ for all $l \in [r+1]$, $\sum_{l \in [r]} i_l = k$, $\sum_{l \in [r+1]} o_l = n-k$ and $\exists l \in [r]$ such that $i_l \geq m$, $\forall r \in [k - m + 1]$. Let $\Kappa_3$ denote the set of all such $k$ sized subsets.
            \item $1 \in \Kappa$ and $n \in \Kappa$. This correspond to sequences of the form $i_1 , o_1 , \hdots , o_{r-1} , i_r$ where $i_l , o_l \geq 1$ for all $l \in [r]$, $\sum_{l \in [r]} i_l = k$, $\sum_{l \in [r]} o_l = n-k$ and $\exists l \in [2:r-1]$ such that $i_l \geq m$ or $x_1 + x_r \geq m$, $\forall r \in [k - m + 1]$. Let $\Kappa_4$ denote the set of all $k$ sized subsets.
        \end{enumerate}
        Now we have $cyc(n , k , m) = |\Kappa_1| + |\Kappa_2| + |\Kappa_3| + |\Kappa_4|$. We will calculate the size of the sets $\Kappa_1, \Kappa_2, \Kappa_3, \Kappa_4$ individually in the following subsections.

        \subsection{Calculation of $\bigl|\Kappa_1\bigr|$}
            Sets in $\Kappa_1$ correspond to positive integer sequences of the form $i_1 , o_1 , \hdots , i_r , o_r$. Here $\sum_{l \in [r]} i_l = k$ and $\sum_{l \in [r]} o_l = n-k$, at least one $i_l \geq m$ and $r$ take all possible values in $[k - m + 1]$.

            Consider $I_j^r$ to be the set of tuples of $r$ positive integers with sum of integers equal to $k$ and the $j^{th}$ integer greater than or equal to $m$, i.e.
            \[
                I_j^r = \{ (i_1 , i_2 , \hdots , i_r) : \sum_{l \in [r]} i_l = k , i_l \geq 1, \forall l \in [r], i_j \geq m \}.
            \]
            For a given $r$, $\cup_{j \in [r]}I_{j}^{r}$ is set of all $r$ length sequences, $(i_1 , \hdots , i_r)$, of positive integers such that $\sum_{l \in [r]} i_l = k$. For all such sequences $i_{1} , \hdots , i_{l}$ there also exist $\binom{n-k-1}{r-1}$ sequences of positive integers $o_{1} , \hdots , o_{r}$ such that $\sum_{l \in [r]} o_l = n -k$. Therefore,
            \[
                |\Kappa_1| = \sum_{r \in [k-m+1]} \binom{n-k-1}{r-1} \bigl|\bigcup_{j \in [r]}I_{j}^{r}\bigr|.
            \]
            From inclusion-exclusion principle we know that
            \[
                \bigl|\bigcup_{j \in [r]}I_{j}^{r}\bigr| = \sum_{l=1}^r(-1)^{l-1}\sum_{1\leq j_1<\cdots<j_l\leq r}\bigl|I_{j_1}\cap\cdots\cap I_{j_l}\bigr| 
            \]
            where,
            \begin{align*}
                &\bigl|I_{j_1}\cap\cdots\cap I_{j_l}\bigr| \\
                &= |\{(i_1 , \hdots , i_r) : \sum_{l \in [r]} i_l = k , i_l \geq 1, \forall l, i_{j_1}, \hdots , i_{j_l} \geq m\}| \\
                &= |\{(i_1 , \hdots , i_r) : \sum_{l \in [r]} i_l = k - l(m-1) , i_l \geq 1, \forall l \in [r]\}| \\
                &= \binom{k - l(m-1) - 1}{r - 1}
            \end{align*}
            which implies,
            \begin{align*}
                &|\Kappa_1| \\
                &= \sum_{r \in [k-m+1]} \binom{n-k-1}{r-1} \bigl|\bigcup_{j \in [r]}I_{j}^{r}\bigr| \\
                &= \sum_{r \in [k-m+1]} \Bigl( \binom{n-k-1}{r-1} \times \\
                &\sum_{l \in [r]} {(-1)}^{l-1} \sum_{1\leq j_1<\cdots<j_l\leq r} \binom{k - l(m-1) - 1}{r-1} \Bigr) \\
                &= \sum_{r \in [k-m+1]} \Bigl( \binom{n-k-1}{r-1} \times \\
                &\sum_{l \in [r]} {(-1)}^{l-1} \binom{r}{l} \binom{k - l(m-1) - 1}{r-1} \Bigr).
            \end{align*}

        \subsection{Calculation of $\bigl|\Kappa_2\bigr|$}
            By the definition of the set $\Kappa_2$ and from sequence of integers $o_1 , i_1 \hdots o_r , i_r$ corresponding to $\Kappa_2$, it is clear that \[ |\Kappa_2| = |\Kappa_1|.
            \]

            \subsection{Calculation of $\bigl|\Kappa_3\bigr|$}
                Here again we see that we need sequence of positive integers $i_1 , \hdots , i_r$ such that $\sum_{l \in [r]} = k$ and $\exists l \in [r]$ for which $i_l \geq m$. We have already calculated this quantity for $|\Kappa_1|$, but for every such sequence of integers, there exist $\binom{n-k-1}{r}$ sequences $o_1 , \hdots , o_{r+1}$ of positive integers such that $\sum_{l \in [r+1]} o_l = n-k$. Therefore,
                \begin{align*}
                    &|\Kappa_3| = \\
                    &\sum_{r = 1}^{k-m+1} \binom{n-k-1}{r} \sum_{l \in [r]} {(-1)}^{l-1} \binom{r}{l} \binom{k - l(m-1) - 1}{r-1}.
                \end{align*}

            \subsection{Calculation of $\bigl|\Kappa_4\bigr|$}
                Consider all sequences of integers $i_1 , o_1 , \hdots , o_{r-1} , i_{r}$ corresponding to $\Kappa_4$ such that $i_l , o_l \geq 1$ for all $l \in [r]$, $\sum_{l \in [r]} i_l = k$, $\sum_{l \in [r]} o_l = n-k$ and $r \geq 2$ and $\exists l \in [r]$ such that $i_l \geq m$ OR $i_1 + i_r \geq m$. $\Kappa_4$ can be partitioned into two disjoint subsets, $\Kappa_{41}$ corresponding to sequences where $i_1 + i_r < m$ and $i_l \geq m$ for at lest one $l \in [2:r-1]$ and $\Kappa_{42}$ corresponding to sequences where $i_1 + i_r \geq m$. Again, $\Kappa_4 = \Kappa_{41} \cup \Kappa_{42}$ and $\Kappa_{41} \cap \Kappa_{42} = \phi$. We will calculate cardinality of both these sets separately
                \subsubsection{$\bigl|\Kappa_{41}\bigr|$}
                    Consider set of all $r$ length positive integer sequences $i_1 \hdots i_r$ such that $i_1 + i_r < m$ and $\sum_{l \in [r]} i_l = k$ and $i_l \geq m$ for some $l \in [2:r-1]$. Note that, for such sequences, $r > 3$. Numbers of such sequences will be 
                    \begin{align*}
                        &\bigl|\{ (i_1 , \hdots , i_r) : \sum_{l \in [r]}i_l = k , i_l \geq 1 , i_1 + i_r < m , \exists l \mbox{ s.t. } i_l \geq m\}\bigr| \\
                        &= \sum_{s=2}^{m-1}(s-1)\bigl|\{(i_2 \hdots i_{r-1}) : \\
                        &\sum_{l = 2}^{r-1} i_l = k - s , i_l \geq 1 , \exists l \mbox{ s.t. } i_l \geq m \}\bigr| \\
                        &= \sum_{s=2}^{m-1}(s-1)\sum_{j = 1}^{r-2} {(-1)}^{j-1} \binom{r-2}{j} \binom{k - s - j(m-1) -1}{r -3}
                    \end{align*}
                    For every such $r$ length sequence, there exist $\binom{n-k-1}{r-2}$ positive integer sequences $o_1 \hdots o_{r-1}$ such that $\sum_{l \in [r-1]} o_{l} = n-k$, we get that
                    \begin{align*}
                        &|\Kappa_{41}| = \\
                        &\sum_{r = 3}^{k-m+1} \Bigl( \binom{n-k-1}{r-2} \sum_{s = 2}^{m-1} (s-1) \times \\
                        &\sum_{j = 1}^{r-2} {(-1)}^{j-1} \binom{r-2}{j} \binom{k - s - j(m-1) -1}{r -3} \Bigr)
                    \end{align*}

                \subsubsection{$\bigl|\Kappa_{42}\bigr|$}
                    Consider set of all $r > 2$ length positive integer sequences $i_1 \hdots i_r$ such that $i_1 + i_r \geq m$ and $\sum_{l \in [r]} i_l = k$. Numbers of such sequences will be 
                    \begin{align*}
                        &\bigl|\{ (i_1 \hdots i_r) : \sum_{l \in [r]} i_l = k , i_1 + i_r \geq m , i_l \geq 1 \} \bigr| \\ 
                        =& \sum_{s = m}^{k-(r-2)} (s-1) \bigl|\{ (i_2 \hdots i_{r-1}) : \sum_{l \in [2:r-1]} i_l = k-s ,i_l \geq 1 \} \bigr| \\ 
                        =& \sum_{s = m}^{k-(r-2)} (s-1) \binom{k-s-1}{r-3}
                    \end{align*}
                    For every such $r$ length sequence, there exist $\binom{n-k-1}{r-2}$ positive integer sequences $o_1 \hdots o_{r-1}$ such that $\sum_{l \in [r-1]} = n-k$, and when $r = 2$, there are $k-1$ possible pairs of positive integers $i_1 , i_2$ which give $i_1 + i_2 = k$, and hence we get  
                    \begin{align*}
                        &|\Kappa_{42}| = \\
                        &\sum_{r = 2}^{k-m+1} \binom{n-k-1}{r-2} \sum_{s = m}^{k-(r-2)} (s-1) \binom{k-s-1}{r-3} + k - 1.
                    \end{align*}
          
           \subsection{Calculation of $cyc(n , k , m) $}
                    Finally, using the calculations of the previous four subsections we have 
        \begin{align*}
            &cyc(n , k , m) = |\Kappa_1| + |\Kappa_2| + |\Kappa_3| + |\Kappa_{41}| + |\Kappa_{42}| \\
            &= \sum_{r = 1}^{k-m+1} \binom{n-k-1}{r-1} \sum_{l \in [r]} {(-1)}^{l-1} \binom{r}{l} \binom{k - l(m-1) - 1}{r-1} \\
            &+ \sum_{r = 1}^{k-m+1} \binom{n-k-1}{r-1} \sum_{l \in [r]} {(-1)}^{l-1} \binom{r}{l} \binom{k - l(m-1) - 1}{r-1} \\
            &+ \sum_{r = 1}^{k-m+1} \binom{n-k-1}{r} \sum_{l \in [r]} {(-1)}^{l-1} \binom{r}{l} \binom{k - l(m-1) - 1}{r-1} \\
            &+ \sum_{r = 3}^{k-m+1} \binom{n-k-1}{r-2} \sum_{s = 2}^{m-1} \Bigl( (s-1) \\
            &\times \sum_{j \in [r-2]} {(-1)}^{j-1} \binom{r-2}{j} \binom{k - s - j(m-1) -1}{r -3} \Bigr)\\
            &+ \sum_{r = 3}^{k-m+1} \binom{n-k-1}{r-2} \sum_{s = m}^{k-(r-2)} (s-1) \binom{k-s-1}{r-3} \\
            &+ k - 1.
        \end{align*}
        Defining $\binom{a}{b} = 0$ if $a < b$ or if $a < 0$ or $b < 0$ the expression above can be simplified to \eqref{cyceqn}. 
        %\begin{align*}
        %    cyc(n , k , m) &= \sum_{r=1}^k \Big(\binom{n-k}{r} + \binom{n-k-1}{r-1}\Big)\sum_{l=1}^r {(-1)}^{l-1}\binom{r}{l}\binom{k - l(m-1) -1}{r-1} \notag \\
        %    &+ \sum_{r=3}^k \binom{n-k-1}{r-2}\Big( \sum_{l=2}^{m-1}(l-1) \sum_{j = 1}^{r-2} {(-1)}^{j-1} \binom{r-2}{j} \binom{k - l - j(m-1) -1}{r-3} \notag \\
        %    &+ \sum_{l=m}^{k} (l-1) \binom{k-l-1}{r-3} \Big) + k-1
        %\end{align*}

    %\bibliographystyle{IEEEtran}
    %\bibliography{reference}

    \section*{Acknowledgement}
    This work was supported partly by the Science and Engineering Research Board (SERB) of Department of Science and Technology (DST), Government of India, through J.C. Bose National Fellowship to B. Sundar Rajan, and by the Ministry of Human Resource Development (MHRD), Government of India, through Prime Minister’s Research Fellowship (PMRF) to Kanishak Vaidya.

    \clearpage
    %\appendix
        %\bibliographystyle{IEEEtran}

\end{document}